# Enhancement of the superconducting transition temperature in Mn-doped CaKFe$_4$As$_4$ processed by the high gas-pressure and high-temperature synthesis method

Manasa Manasa[1], Mohammad Azam[1], Tatiana Zajarniuk[2], Svitlana Stelmakh[1], Tomasz Cetner[1], Andrzej Morawski[1], Shiv J. Singh[1*]

[1]*Institute of High Pressure Physics (IHPP), Polish Academy of Sciences, Sokołowska 29/37, 01-142 Warsaw, Poland*

[2]*Institute of Physics, Polish Academy of Sciences, aleja Lotników 32/46, 02-668 Warsaw, Poland*

*Corresponding author:

 Email: sjs@unipress.waw.pl

https://orcid.org/0000-0001-5769-1787

# Abstract

A series of Mn-doped $CaKFe_4As_4$ samples, $CaK(Fe_{1-x}Mn_x)_4As_4$ with $x$ values of 0, 0.005, 0.01, 0.02, 0.03, 0.04, and 0.05, are synthesized using two distinct routes: conventional synthesis process at ambient pressure (CSP), and high gas-pressure and high-temperature synthesis (HP-HTS) method. Comprehensive characterizations are performed on these samples to investigate their superconducting properties. This study examines the effects of Mn substitution at Fe sites in the FeAs layer on the superconducting properties of the $CaKFe_4As_4$ (1144) material. The HP-HTS process improves the microstructure and phase purity of the parent sample ($x = 0$), resulting in an enhanced superconducting transition temperature ($T_c$). In contrast, Mn doping via the CSP method in $CaKFe_4As_4$ reduces the sample quality and superconducting performance. Notably, the high-pressure synthesis method leads to an increase in the $T_c$ by 3 to 7 K, particularly at low Mn concentrations. While the critical current density ($J_c$) of the parent sample ($x = 0$) shows a significant enhancement under the applied magnetic fields, $J_c$ decreases for Mn-doped $CaKFe_4As_4$ bulks. These results demonstrate that high-pressure synthesis is an effective approach to improve the superconducting properties of Mn-doped 1144 compounds.

# Introduction

    Iron-based superconductors (FBS) [1] exhibit remarkable superconducting properties, including a transition temperature ($T_c$) of up to 58 K, an upper critical field exceeding 60-70 T, and a critical current density reaching up to 100 MA·cm$^{-2}$ [2] [3]. These outstanding parameters make them strong candidates for high-field and practical applications [4] [5]. The high-$T_c$ FBS includes a variety of compounds [6], in which different types of doping: electrons, holes, and isoelectrons, can be introduced, resulting in complex phase diagrams that continue to attract significant research interest [7]. Typically, FBS can be categorized into six families [6]: $RE$FeAsO (1111; $RE$ = rare earth), $Ae$Fe$_2$As$_2$ (122; $Ae$ = Ba, K), Fe$X$ (11; $X$ = Se, Te), $A$FeAs (111; $A$ = Li, Na), $AeA$Fe$_4$As$_4$ (1144; $Ae$ = Ca, $A$ = K) and thick perovskite layered systems such as Sr$_4$V$_2$O$_6$Fe$_2$As$_2$ (42622) and Sr$_4$Sc$_2$O$_6$Fe$_2$P$_2$ (42622). Superconductivity in most of these compounds emerges through suitable chemical doping [3]; for example, F-doped SmFeAsO achieves a notable critical temperature as high as 58 K [8]. In contrast, certain families, such as the 111 and 1144 types, exhibit stoichiometric superconductors. Among them, CaKFe$_4$As$_4$ displays the highest critical temperature ($T_c$) of 35 K [9] [6] [10] [11]. During synthesis, a discrepancy between nominal and actual doping levels often leads to the formation of impurity or secondary phases in doped high-$T_c$ superconductors. However, stoichiometric systems such as the 1144 family inherently avoid these issues and exhibit superconductivity without chemical substitution. Notably, CaKFe$_4$As$_4$ demonstrates a very high critical current density on the order of $10^8$ A/cm² [2]—the highest among all known FBS—making this family particularly attractive for both fundamental studies and potential technological applications, owing to the absence of impurity- or disorder-induced effects associated with doping.

    The layer structure of the 1144 family features alternating layers of alkaline metals and alkaline earth metals positioned between the superconducting FeAs layers. This structure closely resembles that of the 122 family. Interestingly, the 122 phase is not very sensitive to the synthesis temperature, whereas this stoichiometric family is very sensitive to the synthesis temperature, which is around 950°C and a small temperature difference produces the common impurity phases of CaFe$_2$As$_2$ and KFe$_2$As$_2$ [10] [11] [12]. The recent examination of the impurity effects on the superconducting properties of CaKFe$_4$As$_4$ via conventional synthesis process at ambient pressure (CSP), and high gas-pressure and high-temperature synthesis (HP-HTS) method has revealed markedly distinct characteristics in comparison to other FBS families [13], wherein the onset superconducting transition ($T_c^{onset}$) is largely unperturbed by

the existence of prevalent secondary phases such as $CaFe_2As_2$ or $KFe_2As_2$ (122), even when these impurity phases dominate. The observed superconducting transition of the 1144 phase remains consistent at approximately 34-35 K. In other FBS families, the existence of impurity phases generally leads to a swift decrease in their superconducting transition temperature [6] [11] [14]. Previous studies suggest that the CSP method with optimal synthesis conditions (955°C, 6 h) can produce phase-pure 1144 samples [11]. However, these samples generally exhibit weak intergranular connectivity, which severely limits their transport critical current density ($J_c$). The microstructural characteristics of polycrystalline $CaKFe_4As_4$ were investigated using scanning transmission electron microscopy (STEM) and their analysis revealed intrinsic defect structures within the polycrystalline 1144 samples, which play a beneficial role in enhancing the critical current density performance [15]. Consequently, 1144-type polycrystalline samples synthesized by the CSP method typically display low $J_c$ values on the order of $10^3$–$10^4$ A cm$^{-2}$ [11] [15] [16], several orders of magnitude lower than those reported for high-quality 1144 single crystals ($J_c \approx 10^8$ A cm$^{-2}$) [10] [12] [17]. This significant disparity highlights the detrimental effect of weak grain connectivity and the presence of secondary phases in polycrystalline samples. Therefore, it is essential to explore alternative synthesis techniques capable of simultaneously improving grain coupling and phase purity in 1144 bulk materials. Recent studies have demonstrated that high-pressure synthesis can effectively enhance the superconducting properties of $CaKFe_4As_4$ and other iron-based superconductor (FBS) families by improving sample density, structural homogeneity, and crystalline quality [16] [18] [19]. Additionally, spark plasma sintering (SPS) has been employed to densify $CaKFe_4As_4$ bulks, resulting in a substantial improvement in $J_c$ values up to approximately 81 kA/cm² [20]. Furthermore, the spark plasma texturing (SPT) technique was used to fabricate textured $CaKFe_4As_4$ superconducting bulk and enhanced, achieving $J_c$ values of 127 kA cm$^{-2}$ at 4.2 K under self-field and 26 kA cm$^{-2}$ under a magnetic field of 5 T [21]. These finding clearly demonstrate the crucial role of advanced synthesis techniques in optimizing the superconducting performance of 1144-type materials. Importantly, $CaKFe_4As_4$ is a stoichiometric superconductor that exhibits superconductivity without the need for chemical doping [11] [14] [16]. Nevertheless, investigating the effects of selective doping—particularly within the superconducting FeAs layers—remains of significant scientific interest. A recent study of Mn-doped $CaKFe_4As_4$ (1144) single crystals has revealed a significant suppression of superconductivity even at low Mn substitution levels at the iron sites [22] [23]. Essentially, these results align with those observed in other families of FBS, such as Mn-doped SmFeAs(O, F) [24] and Mn-doped $FeSe_{0.5}Te_{0.5}$ [25]. However, to date, no studies have been reported on polycrystalline Mn-doped 1144 samples or their synthesis under high-pressure conditions. Motivated by these findings, the present work explores the high-pressure synthesis and superconducting behavior of Mn-doped CaKFe$_4$As$_4$ bulk materials to elucidate the role of Mn substitution under high-pressure synthesis environments.

In this study, we have prepared a series of Mn-doped CaKFe$_4$As$_4$ bulks using both conventional synthesis process at the ambient pressure and high-pressure methods to investigate the effects Mn substitution at the iron sites within the superconducting FeAs layer. The samples are systematically characterized using various structural and superconducting measurements and compared with previously reported Mn-doped iron-based superconductors [24] [22] [25]. Remarkably, the high-pressure synthesis process enhances the overall sample quality, leading to an improvement in both the superconducting transition temperature ($T_c$) and the critical current density ($J_c$) of the undoped ($x = 0$) 1144 parent compound. Furthermore, Mn doping is found to increase the lattice parameters, confirming the successful incorporation of Mn into the Fe sites [17] [22] [25]. While the CSP-processed samples exhibit a progressive suppression of superconductivity with increasing Mn content, the HP-HTS route significantly improved the superconducting transition by approximately 3–7 K, particularly for doping levels of 2% and 3%.

## Experimental details

Polycrystalline CaK(Fe$_{1-x}$Mn$_x$)$_4$As$_4$ samples are prepared from the initial precursors: calcium powder (purity 99.95%), arsenic chunks (purity 99.999%), potassium (purity 99.9%), manganese (purity 99.3%), and iron powder (purity 99.99%). In the initial step, CaAs is prepared by heating Ca and As at 860 °C for 30 hours, KAs is prepared by heating K and As at 650 °C for 12 hours, and MnAs$_{0.5}$ is prepared by keeping for two step heating of Mn and As at 565 °C for 15 hours and 600 °C for 10 hours. During the preparation of these precursors, the initial elements were placed inside a tantalum (Ta) tube, which was sealed by an ARC melter and subsequently enclosed in an evacuated quartz tube. Furthermore, one more precursor, Fe$_2$As, was synthesized by mixing the elements Fe and As powder, pelletized and sealed in an evacuated quartz tube, which was placed in a furnace at 700 °C for 12 hours. The precursors are mixed according to the stoichiometric formula of CaKFe$_4$As$_4$ and sealed in a Ta tube under an argon atmosphere using an ARC melter. CaKFe$_4$As$_4$ has a very narrow synthesis temperature, as reported in previous studies [11], and even a slight temperature difference of approximately 20°C can lead to the formation of impurity phases. Therefore, sealing the precursors inside a tantalum (Ta) tube using an ARC melter, with extra precautions, was considered necessary to prevent any sudden temperature increases. Otherwise, impurity phases such as $CaFe_2As_2$ can appear alongside the 1144 phase, as reported elsewhere [11] [12]. Hence, we have attempted to prepare multiple 1144 samples by considering the length of the Ta tube and various synthesis temperatures. Furthermore, different synthesis temperatures were utilized to understand their effect on the synthesis process; more details about these samples are provided in our previously published papers [11] [13] [16]. Upon detailed analysis of these samples, we determined the optimal synthesis conditions (955°C for 6 hours) for synthesizing $CaKFe_4As_4$ using the CSP method as an *in-situ* process, taking into account other important parameters. Additionally, we employed the high gas pressure and high-temperature synthesis (HP-HTS) method for the synthesis of the 1144 phase, as outlined in Table 1, which can produce the pressures of up to 1.8 GPa utilizing inert argon gas and a three-zone or one-zone furnace with heating temperatures reaching up to 1700°C, as elaborated elsewhere [18].

In the next step, we have also prepared a series of Mn-doped $CaKFe_4As_4$ ($CaK(Fe_{1-x}Mn_x)_4As_4$; $x = 0, 0.005, 0.01, 0.02, 0.03, 0.04,$ and $0.05$) using a conventional synthesis process (CSP) at ambient pressure and by the HP-HTS method. We utilized starting precursors: CaAs, KAs, $MnAs_{0.5}$, and $Fe_2As$ for the preparation of $CaK(Fe_{1-x}Mn_x)_4As_4$. The optimized synthesis conditions were derived from the aforementioned parent compound $CaKFe_4As_4$. These precursor powders were mixed and sealed in a long Ta tube, which was then placed inside an evacuated quartz tube. In the first step, the samples were heated at 955 °C for 6 hours. The list of these samples is mentioned in Table 1. In the subsequent step, the samples were opened, ground, and then pelletized; they were sealed again in a Ta tube and then in an evacuated quartz tube. This tube was then placed again at 955 °C for only 2 hours. To understand the high-pressure synthesis effect, various Mn-doped $CaKFe_4As_4$ samples were prepared using an *ex-situ* process, where the samples prepared by CSP were sealed in a Ta tube under an inert gas atmosphere and then placed in a high-pressure chamber. We have optimized the high-pressure synthesis conditions for the 11 family, i.e., $FeSe_{0.5}Te_{0.5}$ [18], by preparing many samples, which suggests that 500 MPa for 1 hour is sufficient to obtain high-quality samples. Consequently, these optimized conditions were applied to these 1144 samples. Previous studies [10] [11] have clearly indicated that the prepared 1144 samples are stable up to 600 °C, so we employed 500 °C, 500 MPa, and 1 hour for the preparation of $CaK(Fe_{1-x}Mn_x)_4As_4$ bulk samples via the HP-HTS process to understand the high-pressure synthesis effect, with details about the samples provided in Table 1.

Powder X-ray diffraction (XRD) measurements were carried out using two diffractometers—a Rigaku SmartLab 3 kW and a PANalytical X'Pert PRO—both equipped with filtered Cu-Kα radiation, over a 2θ range of 5°–80°. The use of two instruments ensured cross-verification of the diffraction data and improved the accuracy and reliability of the results. Phase identification and lattice parameter refinement (*a* and *c*) were performed using the ICDD PDF-4+ 2025 database and Rigaku's PDXL analysis software, allowing precise determination of the main tetragonal 1144 phase and any secondary phases present in the Mn-doped CaKFe$_4$As$_4$ samples. Microstructural morphology and compositional analysis were performed on these bulk materials using a Zeiss scanning electron microscope (SEM) attached with a Bruker Quantax 400 energy-dispersive X-ray spectroscopy (EDS) detector by collecting mappings of the component elements and backscattered electron (BSE) images. Electrical resistivity measurements were performed using the conventional four-probe method in a closed-cycle refrigerator (CCR) under zero magnetic field conditions, with data recorded during slow warming. Magnetic measurements, including magnetic susceptibility and hysteresis loop characterization, were conducted using a vibrating sample magnetometer (VSM) integrated into a Quantum Design Physical Property Measurement System (PPMS) under applied magnetic fields up to 9 T.

## Results and discussion

Details of the synthesis conditions for CaK(Fe$_{1-x}$Mn$_x$)$_4$As$_4$ polycrystalline samples are summarized in Table 1. To examine the phase purity and crystal structure, X-ray diffraction (XRD) measurements were performed on all prepared CaK(Fe$_{1-x}$Mn$_x$)$_4$As$_4$ bulks. The diffraction pattern of the selected samples is shown in Figure 1(a), while those of the remaining samples are provided in the supplementary Figure S1. The diffraction patterns confirm that all compositions crystallize in the tetragonal 1144-type structure with space group *I4/mmm* (ThCr$_2$Si$_2$-type structure), consistent with previous reports [9] [10] [11]. The parent CaKFe$_4$As$_4$ prepared by both CSP ($x = 0$) and HP-HTS ($x = 0$_HIP) exhibits a single-phase structure without detectable impurities. At very low level of Mn substitution levels (0.5%), the samples predominantly exhibit the 1144 main phase; however, a small amount of impurity peaks corresponding to CaFe$_2$As$_2$, FeAs and CaFe$_4$As$_3$ phases are also detected in both CSP ($x = 0.005$ and $0.01$) and HP-HTS ($x = 0.005$_HIP and $0.01$_HIP) prepared samples. These impurity phases persist with increasing Mn contents, as shown in Figure 1(a) for $x = 0.02$ and $0.02$_HIP. With further Mn incorporation ($x = 0.03$ and $0.03$_HIP), the relative intensity of these impurity peaks remains nearly unchanged for both synthesis routes. At higher Mn doped samples i.e., $x$ = 0.04 and 0.05, the relative intensity of the impurity phases $CaFe_2As_2$ and FeAs are increased significantly, although the 1144 phase remains the dominant phase (the supplementary Figure S1). Overall, all Mn-doped samples predominantly retain the 1144 main phase; however, the impurity phases such as $CaFe_2As_2$, FeAs, and $CaFe_4As_3$ are consistently detected across all Mn-doped compositions. The fraction of these secondary phases increases noticeably with higher Mn concentrations ($x > 0.03$), indicating a gradual destabilization of the 1144 phase at elevated doping levels.

Based on these patterns, the lattice parameters '$a$', '$c$', and the unit-cell volume '$V$' are extracted and are plotted as a function of Mn content ($x$) in Figure 1(b)-(d) for all $CaK(Fe_{1-x}Mn_x)_4As_4$ polycrystalline samples synthesized via CSP and HP-HTS methods. Since the ionic size of $Mn^{+2}$ (66 pm) is slightly larger than that of $Fe^{+2}$ (63 pm), the substitution of Mn at the Fe sites is expected to expand the lattice parameters. In the HP-HTS processed samples, a slight decrease in lattice parameters '$a$' and '$c$' is observed for low Mn concentrations, i.e., $x = 0.005$ and 0.01, followed by a gradual increase with further Mn-doping. For the samples prepared by CSP, both the lattice parameters '$a$' and '$c$' exhibit a marginal overall increase with increasing Mn content, as illustrated in Figures 1(b) and 1(c). However, a minor reduction in the lattice parameters at the lowest Mn concentration suggests a possible non-uniform distribution of Mn within the FeAs layers. The corresponding variation in the unit-cell volume $V$, presented in Figure 1(d), shows a slight overall increase with Mn substitution, confirming the successful incorporation of Mn into the Fe sites of the 1144 lattice.

Microstructural analyses of the selected $CaK(Fe_{1-x}Mn_x)_4As_4$ bulk samples ($x = 0., 0.02, 0.03, 0\_HIP, 0.02\_HIP,$ and $0.03\_HIP$) is presented in Fig 2(a)-(l). The backscattered electron (BSE) images, depicted in Figure 2(a)-(c) and (g)-(i), illustrate the chemical homogeneity of these samples, where the gray contrast corresponds to the primary 1144 phase, and the black contrast represents pores and, in some regions, $FeAs/Fe_2As$ or other impurity phases, consistent with the XRD results. Overall, BSE images of all these samples look chemical homogeneity. It seems that $x = 0\_HIP$ exhibits the improved densification compared to the sample $x = 0$, indicating the effectiveness of the HP-HTS process in enhancing grain connections. However, Mn doping slightly increases porosity in the samples prepared by the conventional solid-state process. However, this effect is partially mitigated by the HP-HTS method, as seen in Figure 2(g–i). The corresponding elemental mappings, displayed in Figure 2(d–f) and (j–l), show the spatial distributions of Ca, K, Fe, As, and Mn, with distinct colors assigned to each element as indicated. The high-pressure-synthesized sample ($x = 0$_HIP) exhibits nearly homogeneous elemental distribution patterns corresponding to the 1144 phase, confirming uniform incorporation of all constituent elements within the main phase, consistent with the XRD analysis. In contrast, Mn-substitution leads to a gradual deterioration in compositional uniformity and promotes the emergence of minor impurity phases, particularly at higher doping levels ($x = 0.02$ and $0.03$), as observed in both CSP- and HP-HTS-processed samples (Figure 2(e–f) and (k–l)). These observations indicate that Mn doping destabilizes the 1144 phase, promoting the segregation of secondary phases under both synthesis conditions. The microstructural and compositional observations are in good agreement with the XRD results discussed above.

Figure 3 shows the temperature dependence of the resistivity variation of all CaK(Fe$_{1-x}$Mn$_x$)$_4$As$_4$ polycrystalline samples from the normal state to the superconducting region. All samples exhibit almost the same behavior, regardless of whether they were prepared by the CSP or HP-HTS process. The resistivity decreases linearly from room temperature to low temperature and shows superconductivity below 40 K (Figure 3(a)). The normal state resistivity increased for all Mn-doped samples prepared by both CSP and HP-HTS methods compared to their parent samples ($x = 0$). This increase could be attributed to the inhomogeneity of the samples and the presence of pores, as observed from the microstructural analysis, suggesting poor grain connectivity in these Mn-doped samples. The resistivity plot for higher doping concentrations ($x = 0.04$ and $0.05$) is provided in supplementary data (Figure S2), which shows significantly higher resistivity compared to the parent $x = 0$, and there is no clear superconducting transition.

Figure 3(b) illustrates the temperature-dependent resistivity of these samples below the superconducting transition, highlighting both the onset ($T_c^{onset}$) and offset ($T_c^{offset}$) of the superconducting transition temperature. The $x = 0$ sample exhibits a sharp superconducting onset transition ($T_c^{onset}$) at approximately 33.8 K with a narrow transition width of 1 K, whereas the high- pressure-synthesized sample i.e., $x = 0$_HIP shows an enhanced $T_c^{onset}$ of 35.2 K, maintaining the same transition width as that for $x = 0$. Mn-doped samples prepared by the CSP process leads to a rapid suppression of the superconducting transition, with $T_c^{onset}$ values decreasing to 29.2 K and 28.2 K for $x = 0.005$ and $x = 0.01$, respectively. Similarly, the high-pressure processed $x = 0.005$_HIP and $x = 0.01$_HIP samples display comparable or slightly reduced transition temperatures relative to their CSP-synthesized samples ($x = 0.005, 0.01$), as shown in Figure 3(b). With further Mn incorporation ($x = 0.02$ and $0.03$), the superconducting transition is significantly suppressed, with $T_c^{onset}$ values decreasing to 21.8 K and 19 K, respectively, consistent with enhanced magnetic scattering from Mn substitution at the Fe sites [24]. Remarkably, the samples synthesized under high-pressure conditions ($x$ = 0.02_HIP and $x$ = 0.03_HIP) exhibit a notable increase in the transition temperature by approximately 3–7 K compared to their CSP-prepared bulks. Although Mn is generally detrimental to superconductivity, as widely reported for CaK(Fe$_{1-x}$Mn$_x$)$_4$As$_4$ and other iron-based superconductors [17] [22], the present results reveal that high-pressure synthesis can partially counteract this suppression and enhance the superconducting transition, particularly at moderate doping levels (2-3% Mn doping contents). Further investigations are warranted to elucidate the underlying mechanism of Mn incorporation and its interaction with the FeAs layers under high-pressure synthesis conditions.

The temperature dependence of the magnetic susceptibility was measured under an applied magnetic field to confirm the Meissner effect in the CaK(Fe$_{1-x}$Mn$_x$)$_4$As$_4$ samples, as shown in Figure 4(a). Zero-field-cooled (ZFC) and field-cooled (FC) magnetization measurements were performed under a magnetic field of 20 Oe for the $x$ = 0, 0.02, 0_HIP, and 0.02_HIP samples. For comparative analysis, the magnetic susceptibility is normalized to their values at 5 K. The parent compound ($x$ = 0) exhibits a diamagnetic transition at 32.5 K, consistent with the resistivity data and previous reports [9] [11] [16]. The high-pressure-synthesized sample ($x$ = 0_HIP) shows an enhanced and sharper onset transition at 33.9 K, indicating improved electromagnetic homogeneity within the bulk, consistent with earlier observations [11] [16] [15]. The high pressure processed parent $x$ = 0_HIP sample has the sample density around 77% which is slightly higher than the parent $x$ = 0 sample (~65%), as detailed reported elsewhere [16]. Although the magnetic susceptibility does not reach full saturation in the $x$ = 0_HIP sample—likely due to residual porosity—the improvement in diamagnetic response confirms the enhanced intergrain connectivity achieved through the HP-HTS process. This finding suggests that further optimization of the high-pressure synthesis parameters could yield even more denser (up to ~99%) and purer 1144 bulk samples, comparable to those processed by spark plasma sintering (SPS) techniques [20] [21]. Samples synthesized by the HP-HTS method ($x$ = 0_HIP and $x$ = 0.02_HIP) exhibit slightly higher superconducting transition temperatures than their counterparts prepared by the CSP process. In contrast, lightly Mn-doped samples prepared by HP-HTS (≤ 0.01) display nearly identical or marginally lower transition temperatures compared to their ambient-pressure synthesis process, as shown in the supplementary Figures S3(a) and S3(b). The observed $T_c$ values are approximately 28 K and 25 K for $x = 0.005$, 0.005_HIP and $x = 0.01$, 0.01_HIP, respectively, which are consistent with the resistivity results. Furthermore, supplementary Figures S3(c) and S3(d) present the magnetization data for the $x = 0.03$ and $x = 0.03$_HIP samples. The CSP processed sample ($x = 0.03$) exhibits a transition at approximately 10 K, while the high-pressure-synthesized sample ($x = 0.03$_HIP) shows improved phase quality and an enhanced superconducting transition up to 17 K, corroborating the resistivity measurements.

The critical current density ($J_c$) is evaluated for several selected HP-HTS samples, along with the parent sample ($x = 0$) prepared at ambient pressure, as shown in Fig. 4(b). By utilizing the width of the magnetic hysteresis loop, we have calculated the critical current density ($J_c$) using the Bean critical state model, expressed as $J_c = 20\Delta m/Va(1-a/3b)$ [26], where $\Delta m$ represents the difference in magnetization when sweeping the magnetic field up and down, $a$ and $b$ are the short and long edges of the sample (with $a < b$), and $V$ denotes the volume of the sample. Rectangular-shaped samples were employed for the hysteresis loop measurements. For the parent sample $x = 0$, the calculated $J_c$ value reaches $8.5 \times 10^3$ A/cm$^2$ at 0 T, decreasing slightly to $1.8 \times 10^3$ A·cm$^{-2}$ at 9 T, indicating a nearly field-independent behaviour that suggests strong potential for high-field applications. These results are consistent with previous reports on 1144 bulks synthesised by the CSP [15] [11] and SPS methods [20]. Furthermore, 1144 bulks fabricated by the spark plasma texturing (SPT) process have exhibited similar field-dependent behavior, achieving enhanced $J_c$ values of 127 kA·cm$^{-2}$ and 26 kA·cm$^{-2}$ at 4.2 K under self-field and 5 T, respectively [21], demonstrating that $c$-axis texturing can significantly improve the current-carrying performance of 1144 materials. Remarkably, the parent compound synthesized under high pressure ($x = 0$_HIP) exhibits a substantially higher $J_c$ value of $7.5 \times 10^4$ A·cm$^{-2}$ at 5 K and 0 T, representing nearly an order-of-magnitude enhancement compared to the ambient-pressure sample. Despite this improvement, the overall field dependence of $J_c$ remains similar to that of the CSP-processed sample, in agreement with previous studies [10] [11] [16] [22]. This enhancement is likely due to improved sample density, stronger intergrain coupling, and enhanced flux pinning introduced during high-pressure synthesis [27], as similarly observed in MgB$_2$ and other iron-based superconductors [3] [4]. Interestingly, Mn-doped samples synthesized under high pressure display a pronounced suppression of $J_c$ across the entire magnetic field range, although the overall field dependence remains comparable to that of the undoped samples ($x = 0$ and $x = 0$_HIP). The rapid decline in $J_c$ with increasing Mn concentration suggests that Mn incorporation deteriorates flux pinning, likely due to the formation of impurity phases and local inhomogeneity, consistent with previous findings for CaKFe$_4$As$_4$ prepared by different synthesis methods [20] [15] [21].

In Figures 5(a)-(e), the room temperature resistivity ($\rho_{300K}$), onset superconducting transition ($T_c^{onset}$), residual resistivity ratio ($RRR = \rho_{300K}/\rho_{40K}$), transition width ($\Delta T$), and the critical current density ($J_c$) at 0 T are plotted for all these samples prepared by CSP and HP-HTS methods with respect to Mn-doping contents ($x$). The $\rho_{300K}$ value of the samples prepared by CSP shows slight variation with Mn-doping contents, while the high-pressure synthesis of these samples demonstrates non-systematic behavior, as depicted in Figure 5(a). The $T_c^{onset}$ decreases almost linearly with Mn dopants for the samples synthesized by the CSP process. The HP-HTS process for very low amounts of Mn doping ($x = 0.005\_HIP$ and $0.01\_HIP$) yields nearly the same $T_c$ value as that of $x = 0.005$ and $0.01$; however, for slightly higher Mn doping contents, i.e., $x = 0.02\_HIP$ and $0.03\_HIP$, the $T_c$ value has increased by 3–7 K compared to the samples processed by CSP. The $RRR$ value of $x = 0$ is enhanced by high-pressure synthesis, whereas the Mn-doped samples prepared by either the CSP or HP-HTS process show the reduced $RRR$ values (Figure 5(c)), suggesting inhomogeneity of the samples with increased Mn concentration. The transition width ($\Delta T$) of the parent compound $x = 0\_HIP$ is almost the same to that of the sample $x = 0$, however, the sample density of this sample is enhanced by ~15% by HP-HTS process, suggesting the improved intergranular connectivity. However, Mn substitution leads a noticeable broadening of the transition width in both CSP- and HP-HTS prepared samples, as shown in Figure 5(d). This broadening suggests the formation of secondary or impurity phases at the grain boundaries, which becomes more pronounced with increasing Mn content. The obtained $J_c$ values, presented in Figure 5(e), is higher for the parent sample prepared by HP-HTS ($x = 0\_HIP$) compared to the CSP process. Interestingly, all Mn-doped samples produced by the HP-HTS process exhibit a drastic reduction in $J_c$ value, suggesting that Mn doping influences pinning properties or the sample density.

To better understand the influence of Mn doping effects, the obtained lattice parameters ($a$ and $c$) and the unit-lattice volume '$V$' of these samples prepared by the CSP process are compared with those of reported Mn-doped SmFeAs(O,F) (Sm1111) [24] and Mn-doped FeSe$_{0.5}$Te$_{0.5}$ [25], both prepared by the CSP methods (Figures 6(a)-(c)). As no reports are currently available on the high-pressure synthesis of Mn-doped iron-based superconductors, hence, we have not included the lattice parameters of our CaK(Fe$_{1-x}$Mn$_x$)$_4$As$_4$ samples prepared by HP-HTS method in Figures 6(a)-(c). The variation of lattice parameters '$a$' and '$c$' in our samples closely follows the trend reported for Mn-doped SmFeAs(O,F) [24], as depicted in Figures 6(a)-(b). Moreover, the reported lattice parameter '$c$' of Mn-doped FeSe$_{0.5}$Te$_{0.5}$ [25] exhibits a similar dependence, further confirming the consistent structural response to Mn incorporation across different iron-based superconductors like Mn-doped SmFeAs(O,F) [24]. The corresponding lattice volume '$V$' is shown in Figure 6(c) for our samples, Mn-doped Sm1111 [24] and Mn-doped FeSe$_{0.5}$Te$_{0.5}$ [25], which increases systematically with Mn content for all compared systems, consistent with the larger ionic radius of Mn$^{2+}$ relative to Fe$^{2+}$. This expansion of lattice volume strongly supports the successful substitution of Mn at the Fe sites in CaKFe$_4$As$_4$, in agreement with observations for other Mn-doped iron-based superconductors.

Figure 6(d) compares the evolution of the superconducting transition temperature ($T_c$) with Mn concentration for CaK(Fe$_{1-x}$Mn$_x$)$_4$As$_4$ single crystals [22] and polycrystalline samples synthesized by the CSP method. The transition temperature values were determined from both resistivity and magnetic measurements. For comparison, the corresponding data for Mn-doped 1144 samples synthesized via the HP-HTS process are also included in this figure. The CSP-prepared samples exhibit a suppression of $T_c$ with increasing Mn content, following a trend consistent with that reported for Mn-doped 1144 single crystals. Moreover, the absolute $T_c$ values for CSP samples are nearly identical to those of the single crystals across the investigated doping range. Interestingly, the Mn-doped 1144 samples synthesized under high pressure show a similar compositional dependence to those prepared by CSP and in single-crystal form up to the 1% doping level. However, at higher Mn concentrations ($x > 0.01$), the HP-HTS samples display a notable enhancement in $T_c$ by approximately 3–7 K relative to the CSP-prepared ones. This enhancement is particularly evident for the 2% and 3% Mn-doped samples. Although Mn substitution in iron-based superconductors is generally known to strongly suppress superconductivity even at low doping levels [24], our results demonstrate that high-pressure synthesis can partially mitigate this effect and promote higher $T_c$ values up to 7 K in Mn-doped 1144 compounds. Further detailed investigations are therefore essential to elucidate the underlying mechanisms governing Mn incorporation and its influence on the superconducting properties of 1144 and related FBS systems under high-pressure synthesis/growth conditions.

## Conclusions

Mn-doped CaKFe$_4$As$_4$ (CaK(Fe$_{1-x}$Mn$_x$)$_4$As$_4$; $x$ = 0 to 0.05) bulk samples were successfully synthesized using both conventional solid-state processing (i.e., CSP) and high-pressure high-temperature synthesis (i.e., HP-HTS) techniques. Structural analysis revealed a systematic increase in the lattice parameters and unit cell volume with increasing Mn concentration, confirming the successful substitution of Mn at the Fe sites. Microstructural observations, however, indicated that Mn doping levels induced compositional inhomogeneity and secondary phase formation within the samples. The high-pressure synthesis notably improved the superconducting properties of the undoped compound ($x = 0$), enhancing the transition temperature by approximately 1.4 K and increasing the critical current density by nearly one order of magnitude, reflecting better sample densification and intergrain connectivity. Mn substitution in CSP-processed samples led to a pronounced suppression of superconductivity, consistent with previous findings for Mn-doped iron-based superconductors. Remarkably, the high-pressure synthesis of Mn-doped 1144 samples partially counteracted this detrimental effect, leading to a $T_c$ enhancement of 3–7 K for the samples with 2–3% Mn substitution, as confirmed by both resistivity and magnetic measurements. However, the critical current density values of these high-pressure-synthesized samples decreased, suggesting a reduction in effective flux-pinning centers. These findings demonstrate that high-pressure synthesis can be an effective approach for tuning and enhancing the superconducting properties of Mn-doped $CaKFe_4As_4$ and related iron-based superconductors, allowing for the observation of unique characteristics such as the increase of $T_c$ due to Mn doping. Further systematic investigations are required to elucidate the underlying microscopic mechanisms governing the interplay among Mn doping, pressure-induced effects, and superconductivity in the 1144 family and other related iron-based superconducting systems.

## CRediT authorship contribution statement

**Manasa Manasa:** Writing – review & editing, Investigation, Writing – original draft, Investigation, Formal analysis, Data curation. **Mohammad Azam:** Writing – review & editing, Data curation. **Tatiana Zajarniuk:** Investigation, Data curation. **Svitlana Stelmakh:** Formal analysis, Investigation, Data curation. **Tomasz Cetner:** Writing – review & editing, Data curation. **Andrzej Morawski:** Data curation. **Shiv J. Singh:** Writing – review & editing, Writing – original draft, Visualization, Validation, Supervision, Software, Resources, Methodology, Investigation, Funding acquisition, Formal analysis, Conceptualization.

## Declaration of competing interest

The authors declare that they have no known competing financial interests or personal relationships that could have appeared to influence the work reported in this paper.

## Data availability

## Acknowledgments:

The work was funded by SONATA-BIS 11 project (Registration number: 2021/42/E/ST5/00262) sponsored by National Science Centre (NCN), Poland. SJS acknowledges financial support from National Science Centre (NCN), Poland through research Project number: 2021/42/E/ST5/00262.

**Table 1:** Synthesis conditions and corresponding sample codes for Mn-doped CaKFe$_4$As$_4$ (CaK(Fe$_{1-x}$Mn$_x$)$_4$As$_4$) polycrystalline samples with $x$ = 0, 0.005, 0.01, 0.02, 0.03, 0.04, and 0.05, prepared using the CSP at ambient pressure and the HP-HTS methods.

| Sample Code ($x$) | Synthesis Conditions | Synthesis Process |
|---|---|---|
| 0 | *First step:* heated at 955 °C, 6 h, 0 MPa (Sealed inside Ta-tube) ↓ *Second step:* heated at 955 °C, 2 h, 0 MPa (Sealed inside Ta-tube) | *In-situ* |
| 0.005 | | *In-situ* |
| 0.01 | | *In-situ* |
| 0.02 | | *In-situ* |
| 0.03 | | *In-situ* |
| 0.04 | | *In-situ* |
| 0.05 | | *In-situ* |
| 0_HIP | 500 °C, 1 h, 500 MPa (Sealed inside Ta tube) | *Ex-situ* |
| 0.005_HIP | | *Ex-situ* |
| 0.01_HIP | | *Ex-situ* |
| 0.02_HIP | | *Ex-situ* |
| 0.03_HIP | | *Ex-situ* |

**Figure 1:** **(a)** Powder XRD pattern of CaK(Fe$_{1-x}$Mn$_x$)$_4$As$_4$ polycrystalline samples with $x$ = 0, 0_HIP, 0.02, 0.02_HIP, 0.03 and 0.03_HIP is presented. The variation of **(b)** Lattice constant '$a$', **(c)** Lattice constant '$c$', and **(d)** Unit cell volume '$V$' is shown for all CaK(Fe$_{1-x}$Mn$_x$)$_4$As$_4$ samples synthesized using the CSP under ambient pressure and the HP-HTS method with the Mn doping contents ($x$).

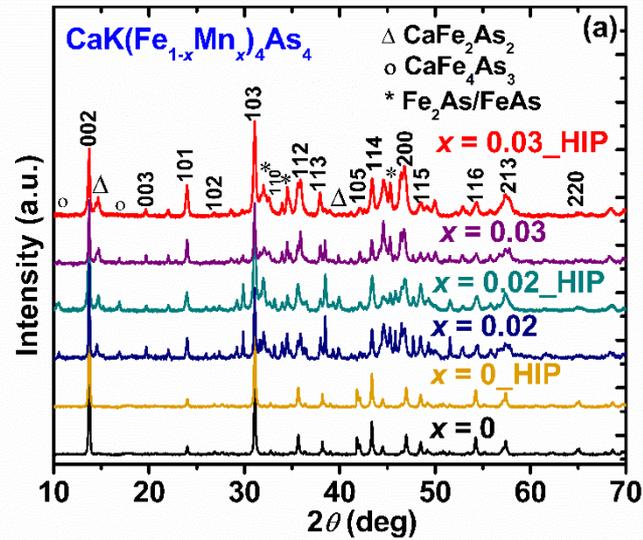

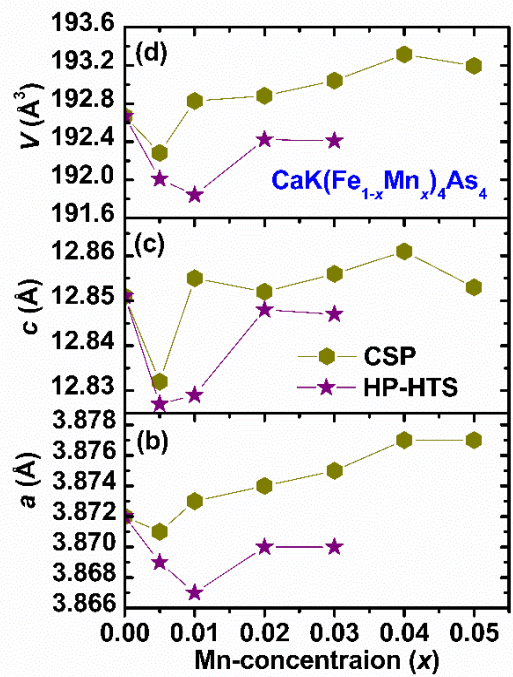

**Figure 2:** Backscattered electron (BSE) images and the corresponding elemental mapping for CaK(Fe$_{1-x}$Mn$_x$)$_4$As$_4$ polycrystal samples: **(a)**, **(d)** $x = 0$, **(b)**, **(e)** $x = 0.02$ **(c)**, **(f)** ) $x = 0.03$, **(g)**, **(j)** 0_HIP **(h)**, **(k)** $x = 0.02$_HIP and **(i)**, **(l)** $x = 0.03$_HIP. The light gray and black contrasts in the BSE images correspond to the 1144 phase and pores, respectively. Occasionally, the black contrast may represent FeAs/Fe$_2$As impurity phases in the parent samples, and other secondary phases in the Mn doped CaKFe$_4$As$_4$ samples.

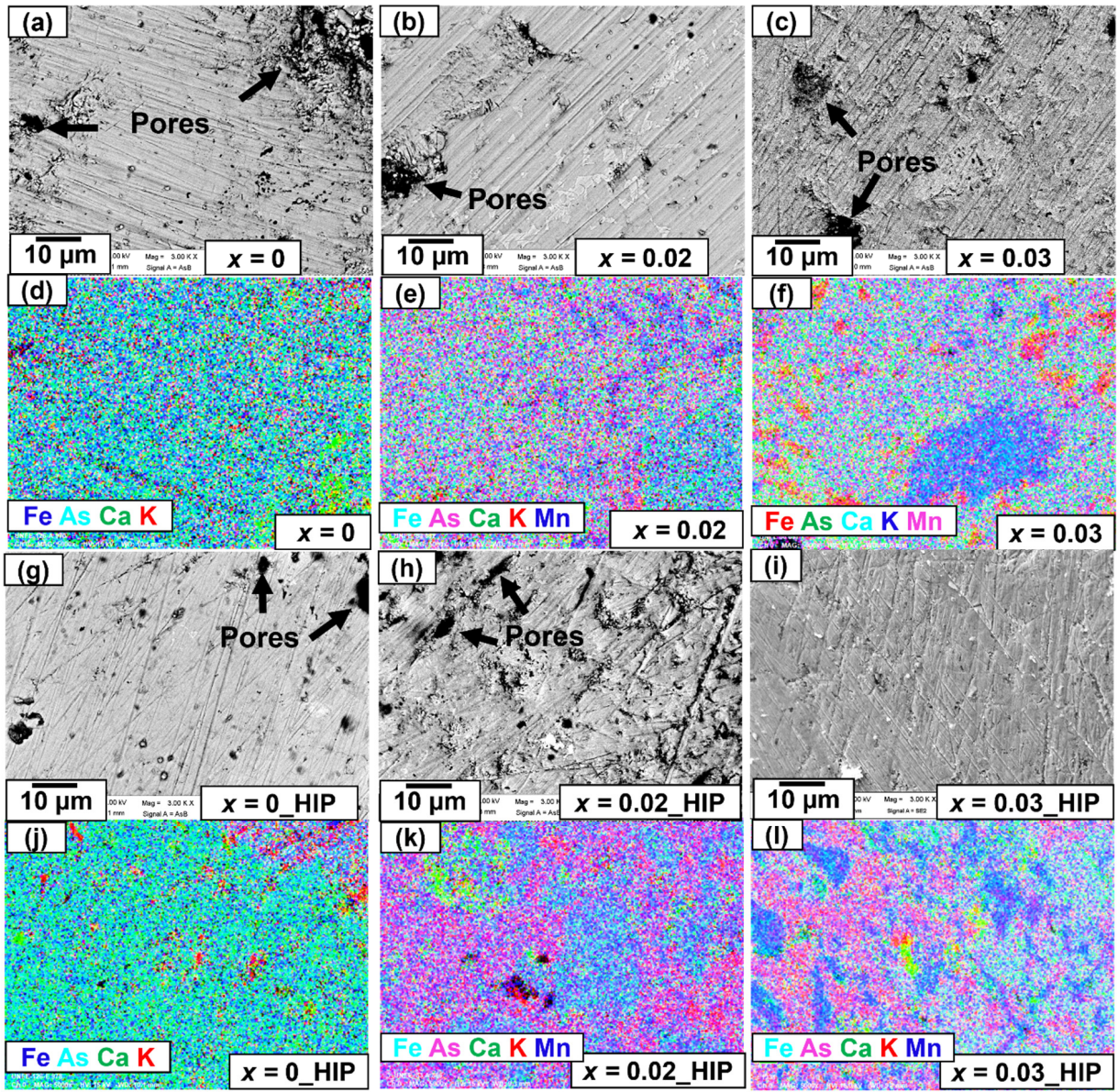

**Figure 3:** **(a)** The temperature dependence of electrical resistivity ($\rho$) up to room temperature, and **(b)** the low-temperature variation of resistivity up to 40 K for CaK(Fe$_{1-x}$Mn$_x$)$_4$As$_4$ polycrystalline samples prepared by both the CSP and HP-HTS methods.

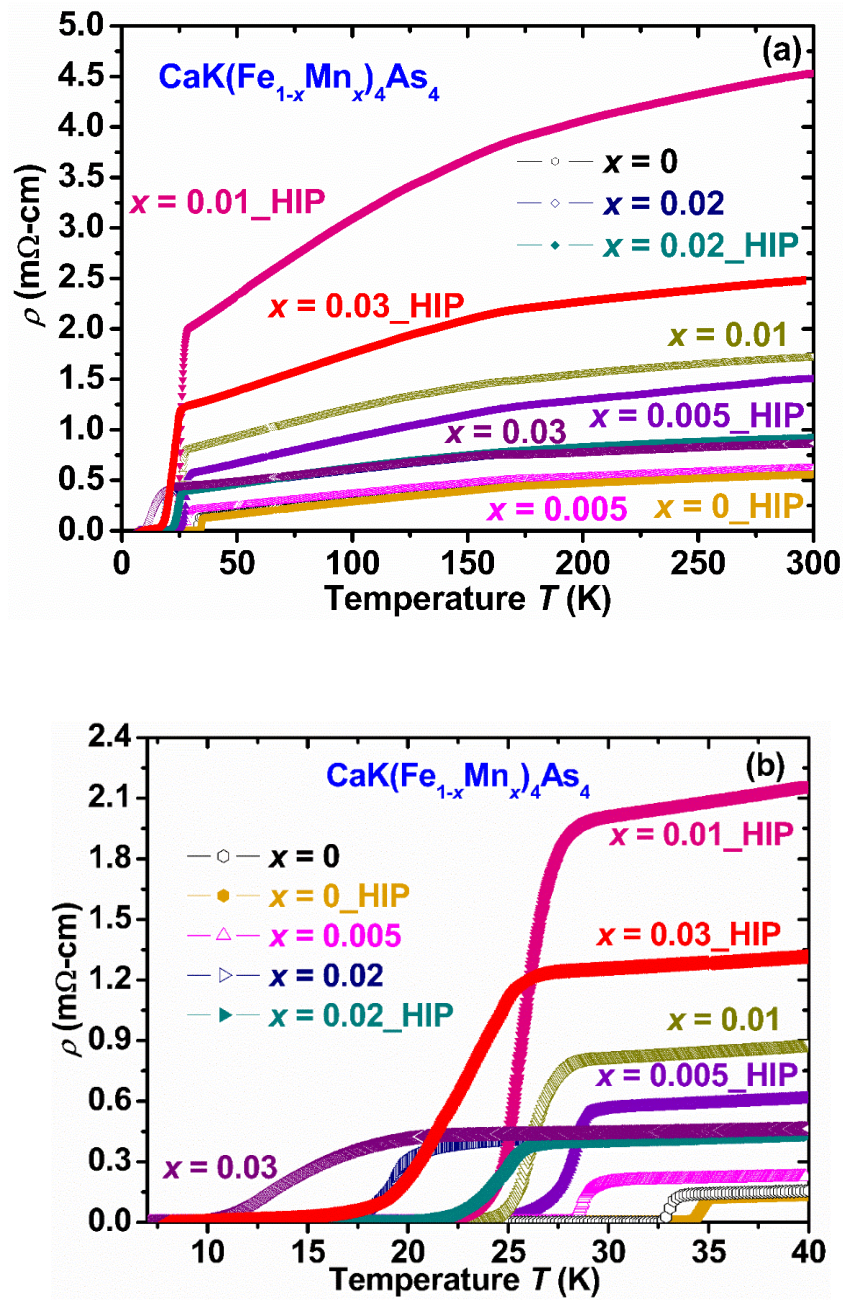

**Figure 4: (a)** The temperature dependence of the normalized magnetic susceptibility ($4\pi M/H$) measured in zero-field-cooled (ZFC) and field-cooled (FC) modes for CaK(Fe$_{1-x}$Mn$_x$)$_4$As$_4$ bulk samples with $x = 0$, $x = 0$_HIP, $x = 0.02$, $x = 0.02$_HIP under an applied magnetic field of 20 Oe. **(b)** The magnetic field variation of the critical current density ($J_c$) at 5 K up to the magnetic field of 9 T for the samples with $x = 0$, $x = 0$_HIP, $x = 0.005$_HIP, $x = 0.01$_HIP $x = 0.02$_HIP and $x = 0.03$_HIP.

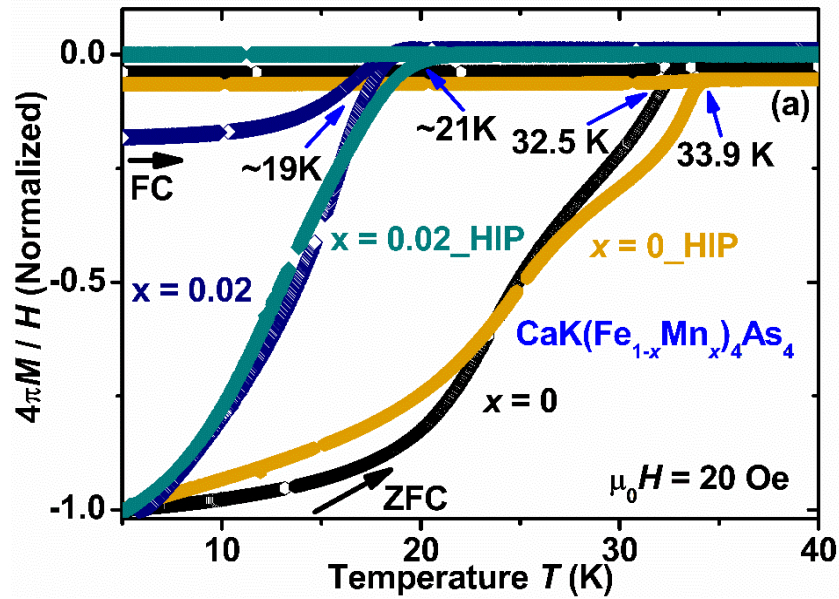

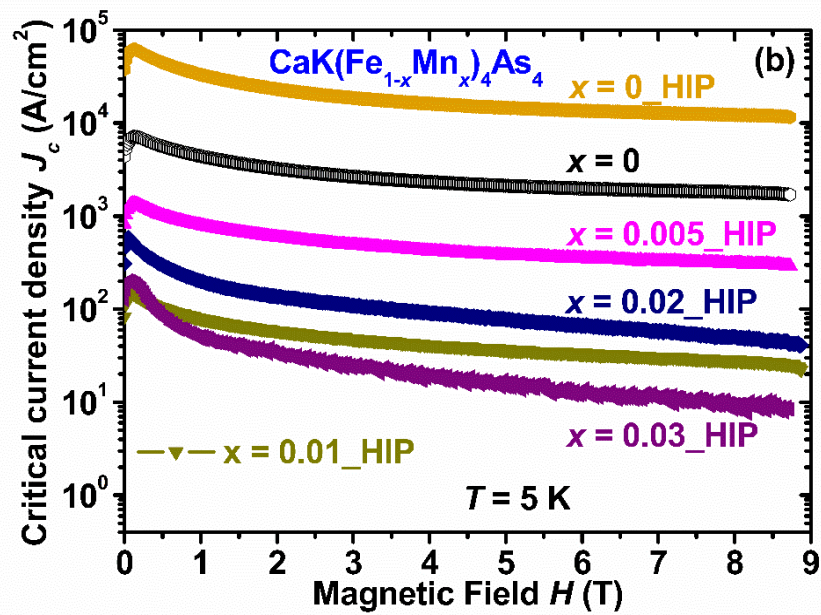

**Figure 5**: The variation of **(a)** the room temperature resistivity ($\rho_{300K}$), **(b)** the superconducting onset transition temperature $T_c^{onset}$, **(c)** the residual resistivity ratio ($RRR = \rho_{300K}/\rho_{40K}$), **(d)** transition width ($\Delta T = T_c^{onset} - T_c^{offset}$), and **(e)** the critical current density ($J_c$) at 0 T at 5 K as a function of Mn-concentration ($x$) are presented for CaK(Fe$_{1-x}$Mn$_x$)$_4$As$_4$ polycrystalline samples prepared by the CSP at ambient pressure and the HP-HTS method.

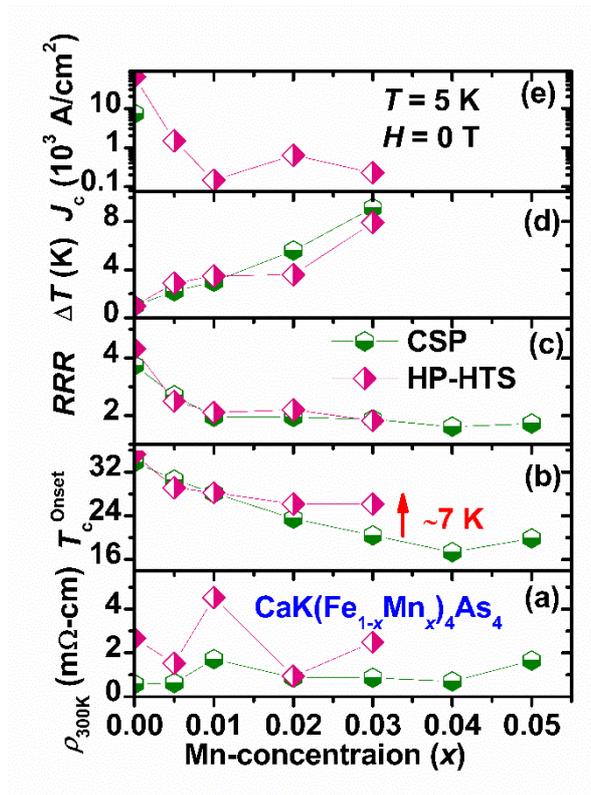

**Figure 6:** Normalized graph of **(a)** the lattice constant '$a$', **(b)** the lattice constant '$c$', and **(c)** the unit cell volume '$V$' as a function of Mn doping content ($x$) are presented for all CaK(Fe$_{1-x}$Mn$_x$)$_4$As$_4$ samples synthesized by the CSP method at ambient pressure, compared with previously reported data for Mn-doped SmFeAs(O,F) (denoted as "1111" in the figure) [24] and Mn-doped FeSe$_{0.5}$Te$_{0.5}$ (denoted as "11" in the figure) [25]. **(d)** A comparative summary of the superconducting transition temperatures ($T_c$) of CaK(Fe$_{1-x}$Mn$_x$)$_4$As$_4$ bulks prepared by CSP at ambient pressure and HP-HTS methods, together with reported Mn-doped CaKFe$_4$As$_4$ single-crystals grown by the conventional method at ambient pressure [22], is shown with Mn concentration ($x$). The transition temperature values are determined from both the resistivity and magnetic measurements.

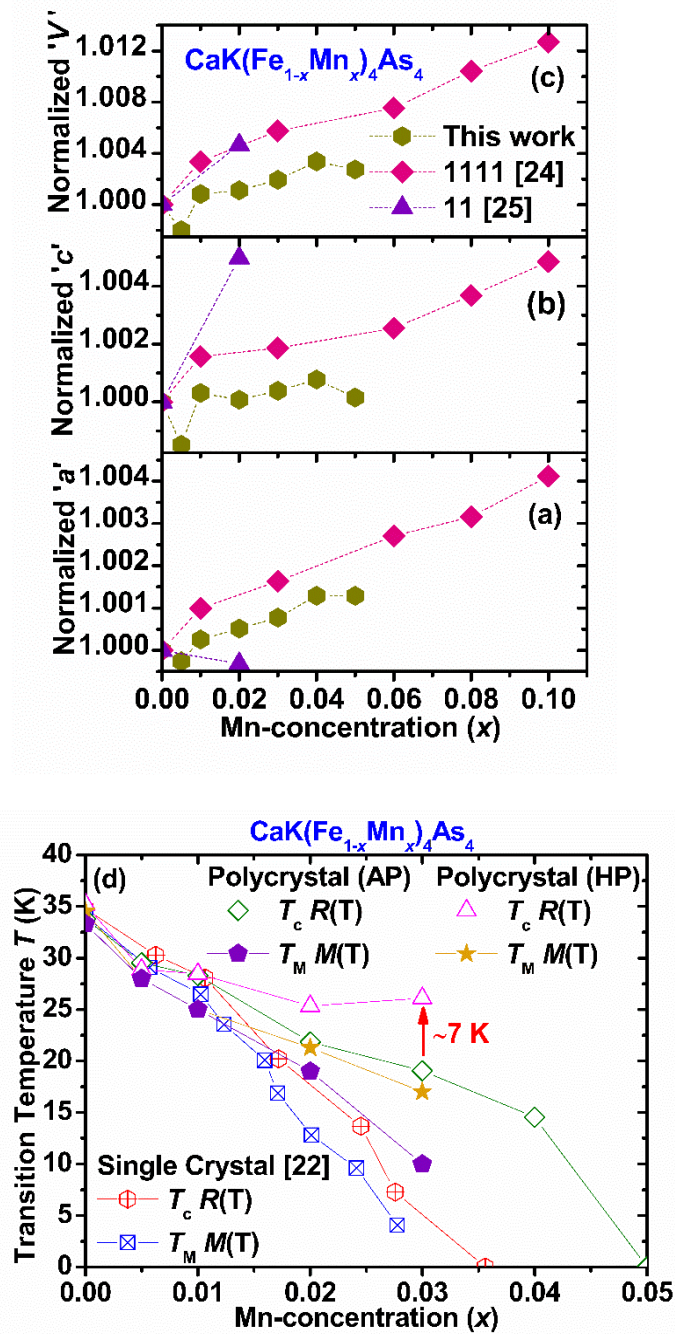